\documentclass[11pt]{article}
\usepackage{graphicx,moriond,epsfig,amsmath}

\def\Journal#1#2#3#4{{#1} {\bf #2}, #3 (#4)}

\def\APP{\em Astroparticle Physics}
\def\arxiv{arXiv:}
\def\JHEP{JHEP}
\def\MNRAS{\em Mon. Not. Roy. Astron. Soc.}
\def\NIMA{{\em Nucl. Instrum. Methods} A}
\def\PLB{{\em Phys. Lett.}  B}
\def\PR{\em Phys. Rep.}
\def\PRD{{\em Phys. Rev.} D}

\newcommand{\comment}[1]{}
\newcommand{\kms}{km\,s$^{-1}$}
\newcommand{\kgd}{kg$\cdot$d}

\begin{document}
\vspace*{4cm}
\title{LATEST RESULTS OF THE EDELWEISS-II EXPERIMENT}

\author{ A.S. TORRENT\'O-COELLO }

\address{IRFU/SPP, CEA-Saclay, 91191 Gif-sur-Yvette, France}

\maketitle\abstracts{
The EDELWEISS-II collaboration has performed a direct search for WIMP dark matter with an array of ten 400-g heat-and-ionization cryogenic detectors equipped with Inter-Digit electrodes for the rejection of near-surface events. Results from one year of continuous operation at the Laboratoire Souterrain de Modane will be presented. A sensitivity to the spin-independent WIMP-nucleon elastic cross-section of 4.4$\times$10$^{-8}$~pb was achieved using a 384~\kgd~effective exposure. We also interpret the results in the inelastic scattering scenario, excluding the DAMA allowed region for WIMP masses greater than 90~GeV for a mass splitting of 120~keV. The results obtained demonstrate the excellent background rejection capabilities of these simple and robust detectors in an actual WIMP search experiment. Some first results with 800-g detectors will be also presented together with the prospects for this experiment and the ton-scale EURECA project.}

\section{The WIMP search and the EDELWEISS-II setup}
The existence of Weakly Interacting Massive Particles (WIMPs) is a likely explanation for the various observations of a non-luminous, non-baryonic matter component from the largest scales of the Universe to galactic scales \cite{Ber2005}. The WIMP search tests the hypothesis of the Milky Way being surrounded by a dark matter halo constituted by these particles. WIMPs would scatter off ordinary target nuclei on Earth yielding a low-energy deposit of the order of 10\nobreakdash--100~keV. The expected interaction rate is very low, and it is currently constrained at the level of $<$1~event/kg/year. The detection of such small signals is technically difficult, requiring an ultralow radioactivity environment and detectors with a low energy threshold which could actively reject the residual backgrounds.\\	

EDELWEISS is an experiment aimed at the direct detection of dark matter with ultrapure Ge bolometers. The EDELWEISS-II setup is installed in the Laboratoire Souterrain de Modane (LSM), the deepest underground laboratory in Europe. The 4800~m~water-equivalent of rock above the experimental cavity reduces the cosmic muon flux to 4~$\mu$/m$^2$/day. The fast neutron flux from the rock has been measured to be 10$^{-6}$~n/cm$^2$/s.\\

The 400-g Ge bolometers are operated at 20~mK in a dilution refrigerator which can be controlled remotely. The bolometers are protected against the relevant background contributions, i.e.~gamma and beta radioactivity, cosmic muons and neutrons, by means of passive shields and active rejection.
A 20-cm thick lead shield, with an inner part made up with roman lead, surrounds the detectors to attenuate the external $\gamma$ radioactivity. It follows a 50-cm polyethylene shield which attenuates the fast neutron flux. An active muon veto system made of plastic scintillators covers the whole setup with a 98\% geometrical efficiency. The setup is installed in a clean room with deradonised air to minimise the beta contamination from Rn. Additional background monitoring is achieved by a Rn detector near the cryostat, a $^3$He-gas detector inside the shielding (thermal neutron flux), and a Gd-loaded liquid scintillator outside the shielding (muon-induced neutrons). Radioactive sources are available for gamma ($^{133}$Ba), beta (Rn), and neutron (AmBe) calibration of the detectors.\\

The bolometers are instrumented to perform a dual heat and ionisation measurement of the signals arising from particle interactions in order to discriminate $\gamma$-induced electronic recoils from potential WIMP-induced nuclear recoils.
The heat is measured with NTD \footnote{Neutron Transmutation Doped.} sensors glued on the surface of each detector. The ionisation signal is measured with a set of Al electrodes deposited on the bolometer surface, which are polarised at a few V/cm. The electronic/nuclear recoil discrimination is performed through the quantity called "ionisation yield" ($Q$) which is the ratio between the ionisation and recoil energies. By definition, $Q$=1 for electronic recoils and it is a factor $\sim$3 lower for nuclear recoils in Ge in the energy region of interest. This method allows to perform an event-by-event rejection of the bulk of $\gamma$ radioactivity. However, when the interaction takes place near the detector surface the charge collection is incomplete, and the value of $Q$ for electronic recoils can be at the level of that for nuclear recoils, leading to a mis-identification of these events. These surface events are mainly generated by local $\beta$ radioactivity from $^{210}$Pb, a daughter of Rn present in air which is deposited on all material surfaces.\\

\begin{figure}[h!]
\begin{center}
 \begin{minipage}{0.49\linewidth}
       \includegraphics[width=\linewidth]{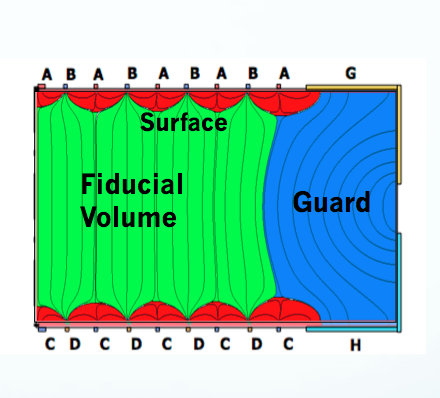}
\end{minipage}
\hfill   
\begin{minipage}{0.49\linewidth}
      \includegraphics[width=\linewidth]{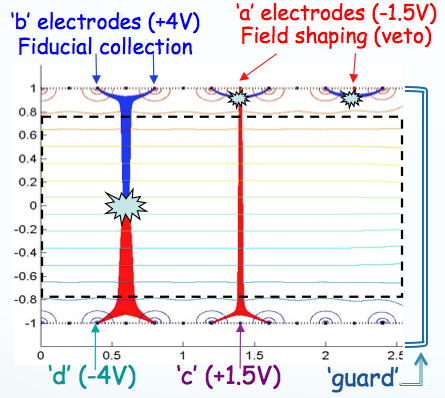}
\end{minipage}
\caption{Left: Topology of the electric field created by interleaved electrodes. Right: Functioning principle of Inter-Digit detectors for a fiducial event (left) and two surface events (center, right). The lines represent the electrostatic equipotentials.}
\label{fig:ID400}
\end{center}
\end{figure}

To get rid of this intrinsic background, a new generation of detectors so-called "ID" (Inter-Digit) has been developped. The functioning principle of these detectors is shown in figure~\ref{fig:ID400}. The electrodes are concentric circles connected alternatively, forming 2 interleaved sets: one for ionisation charge collection and one for electric field shaping ("veto"). The electric field created is vertical in the bulk of the bolometer and near-horizontal next to the top and bottom surfaces. Two more electrodes covering the lateral surfaces ("guard") limit the detection volume to the bulk.

An interaction in the fiducial volume reaches both collecting electrodes with a perfect charge balance, leaving no signal in any of the veto or guard electrodes. An interaction on the surface is seen by the veto or guard electrodes, and the signal is asymmetrically collected by the collecting electrodes. This redundance on the  identification of surface events provides a very high rejection efficiency ($\sim$10$^{-5}$) for gammas and betas \cite{Bro2009}.\\

\section{WIMP-search analysis and results} \label{sec:WIMP search}
A WIMP search was carried out using ten 400-g ID detectors installed in the EDELWEISS-II setup from April 2009 to May 2010. Here we present the analysis of the full data set, which also includes the data from the validation run of the ID technology in 2008. The overall exposure doubles the one in the 6-month WIMP-search published in \cite{EDW2010}.\\

During the whole acquisition period, the cryogenic conditions were maintained stable at $\sim$18~mK without any major interruption. Most of the time was devoted to WIMP search (325~days), and a small fraction to gamma and neutron calibrations (10.1 and 6.4~days, respectively). All heat sensors and 90\% of the ionisation channels were operational. The redundancy in the background rejection allowed to use all the detectors for WIMP search. An online trigger kept the trigger rate below a fraction of Hz.\\

The data was analysed using two independent pipelines which yield consistent results. An optimal filtering algorithm allowed processing the signals accordingly to the changing noise conditions. The average baseline resolutions of heat and fiducial ionisation channels were of $\sim$1.2~keV FWMH and $\sim$0.9~keV FWHM, respectively. Noisy periods were automatically discarded on a baseline-measurement basis with 80\% efficiency. A $\chi^2$ cut was used to reject misreconstructed events. The WIMPs were searched among the fiducial events, using the ionisation yield to discriminate gamma-rays with a 99.99\% rejection efficiency. Tagging of coincident events in bolometers and the muon veto allowed to reject neutron-induced recoils. Finally, a WIMP-search energy threshold was set a priori to 20~keV, so that the search efficiency is independent of the energy. After all cuts, the effective exposure obtained is 384~\kgd. This analysis procedure is identical to the one previously used in \cite{EDW2010}.\\

\begin{figure}[h!]
\begin{center}
 \begin{minipage}[b]{0.49\linewidth}
       \includegraphics[width=\linewidth]{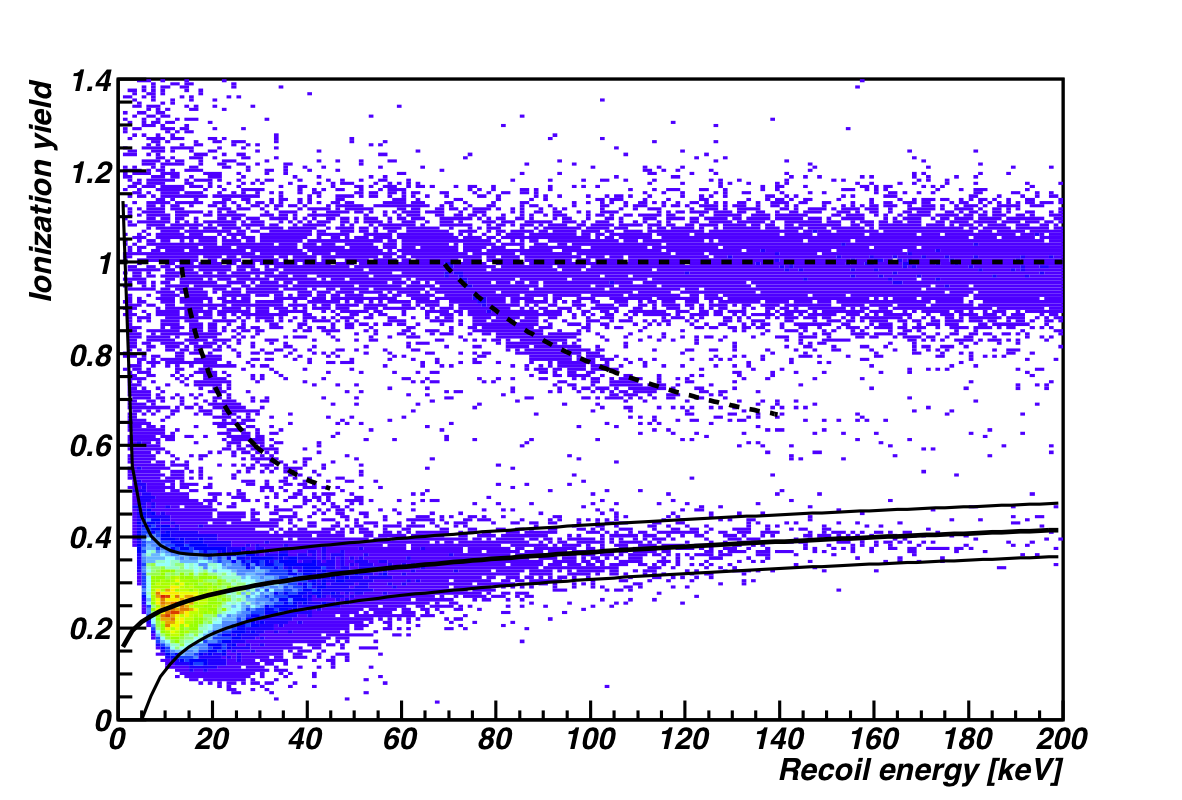}
\end{minipage}   
\begin{minipage}[b]{0.49\linewidth}
      \includegraphics[width=\linewidth]{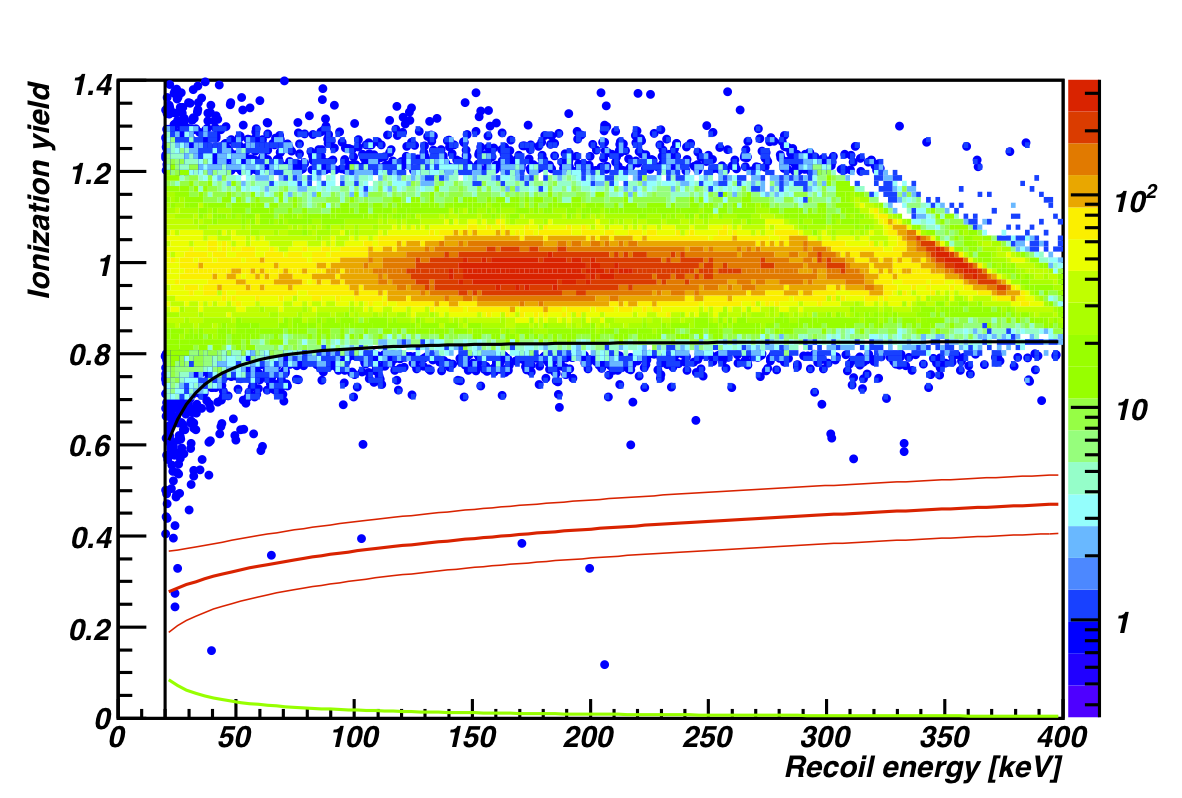}
\end{minipage}
\caption{Ionisation yield vs.~recoil energy. Left: Neutron calibration with an AmBe source. The electronic recoil nominal value $Q$=1, the 90\%-acceptance nuclear recoil band (bottom) and the lines arising from inelastic neutron scattering at 13.26~keV and 68.75~keV (dashed) are highlighted. Right: Gamma calibration with a $^{133}$Ba source. The lines superimposed represent the average 99.99\% rejection line for electron recoils (black), the 90\%-acceptance nuclear recoil band (red) and the typical ionisation threshold (green).}
\label{fig:neutron gamma calibration}
\end{center}
\end{figure}

In figure~\ref{fig:neutron gamma calibration} we show the ionisation yield vs.~recoil energy obtained in the neutron and gamma calibrations. In the neutron calibration, the region with 90\% acceptance for nuclear recoils, which is the one used in the WIMP search, is well described by the parametrisation of \cite{Mar2004} using the measured resolutions of the heat and ionisation signals. From the $\gamma$-ray calibration we obtain a $\gamma$ rejection factor of $\sim$3$\times$10$^{-5}$. The origin of the six events leaking into the nuclear recoil band is being investigated. It has already been verified that they are not related to a specific time period, bolometers with missing electrodes or electrodes with bad resolution.\\

The ionisation yield plot obtained in the WIMP search after fiducial cuts is shown in figure~\ref{fig:qplot_wimp, limit_idm0}~(left). Five events are found in the nuclear recoil band: four of them have energies between 20 and 23~keV, and one has 172~keV. All of them are well-reconstructed events which lie well above the noise level of the detectors. Background studies are ongoing to fully understand their origin. Upper limits may be derived from the known residual gamma, beta and neutron backgrounds, using calibration data, material radioactivity measurements and Monte Carlo simulation of the detectors. Overall, less than 3 events (90\%~CL) from known origin are expected in this WIMP search.\\

The spin-independent cross-section upper limit for WIMP-nucleon elastic scattering is calculated from the presented data following the optimum interval method described in \cite{Ye2001}. A standard halo model is considered, and the calculation of the differential event rate is performed according to the analytical solution proposed by \cite{SFG2006}, with the following values for the relevant parameters: $\rho_{\chi}$=0.3~GeV$c^{-2}$cm$^{-3}$ (local dark matter density), $v_0$=220~\kms (dark matter Maxwellian velocity dispersion),  $v_{\rm earth}$=235~\kms~(average Earth velocity) and a recent estimation of the galactic escape velocity, $v_{\rm esc}$=544~\kms~\cite{RAVE2007}. We include the effect of a detector finite recoil energy resolution of 1.5~keV. The 90\%~CL limit obtained vs.~the WIMP mass is shown in figure~\ref{fig:qplot_wimp, limit_idm0}~(right). The best sensitivity obtained by EDELWEISS-II is 4.4$\times$10$^{-8}$~pb at $M_{\chi}$=85~GeV, which is more than twice as constraining than the one obtained with six months of data \cite{EDW2010}.\\

\begin{figure}[h!]
\begin{center}
 \begin{minipage}{0.49\linewidth}
       \vspace{0pt} \includegraphics[width=\linewidth]{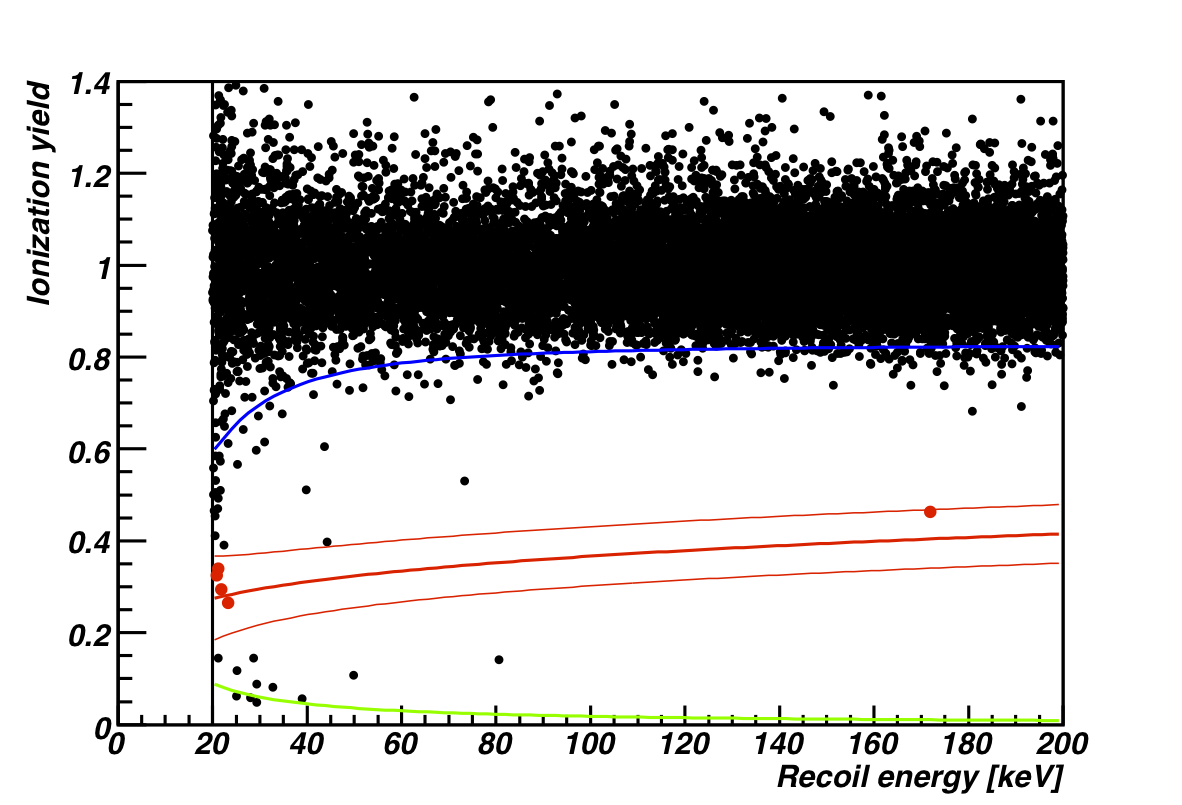}
\end{minipage}
\begin{minipage}{0.49\linewidth}
      \vspace{10pt} \includegraphics[width=0.68\linewidth,height=\linewidth,angle=90]{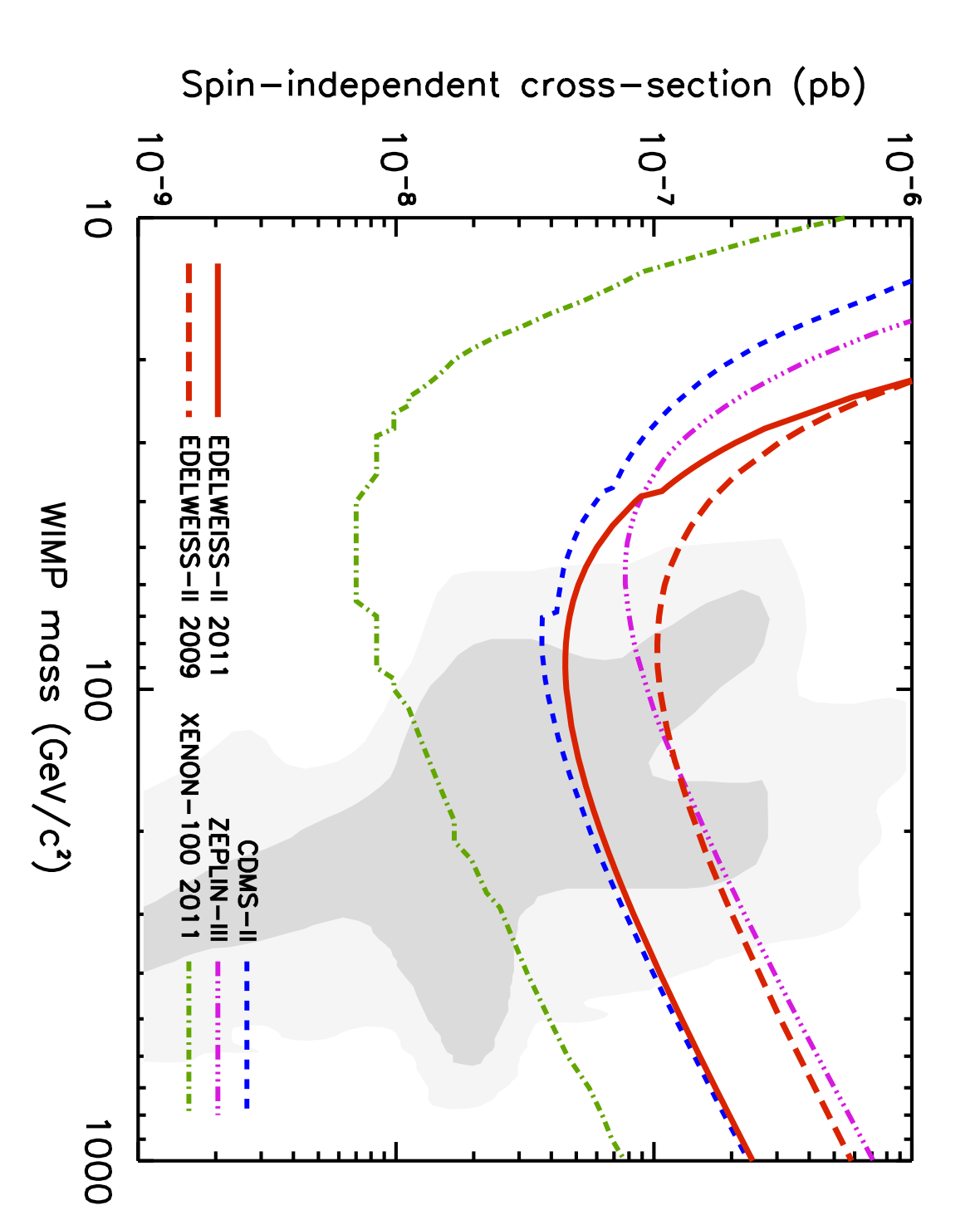}
\end{minipage}
\caption{Left: Ionisation yield vs.~recoil energy obtained in the WIMP search after fiducial cuts. Highlighted in red, the five events found in the nuclear recoil band which are retained as WIMP candidates. Right: WIMP-nucleon spin-independent elastic cross-section upper limit at 90\% CL vs.~WIMP mass obtained in this analysis (solid red) and recent results from other experiments. The shaded area corresponds to the 68\% and 98\% probability regions of the cMSSM scan from \cite{Ros2007}.}
\label{fig:qplot_wimp, limit_idm0}
\end{center}
\end{figure}

\section{Inelastic scattering scenario} \label{sec:inelastic scattering}
The inelastic dark matter scenario has been proposed to reconcile the dark matter modulation signal claimed by DAMA/LIBRA and the null detection in all the other direct detection experiments \cite{SW2001}.\\

In this scenario, the WIMP has a ground state and an excited state $\delta$$\sim$100~keV heavier than the previous one. The WIMP-nucleus scattering would occur through a transition to the excited state ($\chi + \mathcal{N} \to \chi^* + \mathcal{N}$), with the elastic scattering being highly suppressed.\\

The kinematics of the scattering process in the inelastic scenario differs from the elastic one, as the kinetic energy needed for a WIMP to scatter off nuclei is higher in the former case. For the direct detection experiments this translates into a higher minimum velocity to deposit a recoil energy $E_{\rm R}$ in the detector
\begin{equation} \label{eq:vmin}
v_{\rm min}=\frac{1}{c^2}\sqrt{\frac{1}{2 m E_{\rm R}}}\left( \frac{m E_{\rm R}}{\mu} + \delta \right)
\end{equation}
where $m$ is the mass of the target nucleus and $\mu$ is the reduced mass of the WIMP-target nucleus system. The increase of $v_{\rm min}$ results in a dramatic modification of the expected WIMP spectrum, as only the high end of the WIMP halo velocity distribution will contribute to the signal: the event rate is globally reduced and suppressed at low recoil energies, and the modulation signal is enhanced. Moreover, for a given $E_{\rm R}$ and $\delta$, heavier targets will be more sensitive than lighter ones.\\

In both elastic and inelastic scenarios, the differential event rate can be calculated from
\begin{equation} \label{eq: WIMP rate}
\frac{dR}{dE}=\frac{\rho_{\chi}}{2 M_{\chi} \mu^2} \sigma_0  F^2(q) \int_{v_{\rm min}}^{v_{\rm max}} \frac{f(\bf{v})}{v} d^3 v
\end{equation}
where $\rho_{\chi}$ is the local WIMP density, $M_{\chi}$ is the WIMP mass, $\sigma_0$ is the zero-momentum WIMP-nucleus cross-section and $F^2(q)$ is the Helm nuclear form factor for momentum transfer\linebreak
\(q=\sqrt{2m E_R}\), which is described in \cite{LS1996}. We consider a standard WIMP halo model with a Maxwellian velocity distribution characterised by a velocity dispersion $\sigma_v=\sqrt{2/3}v_0$ which is truncated at a galactic escape velocity $v_{\rm esc}$. The integral of the inverse mean of the velocity distribution is done from $v_{\rm min}$ to $v_{\rm max}$=$v_{\rm esc}$+$v_{\rm earth}$, the maximum velocity a WIMP can have to interact with a target nucleus (otherwise it would escape the Galaxy).\\

\begin{figure}[h!]
\begin{center}
        \includegraphics[height=0.7\linewidth,angle=90]{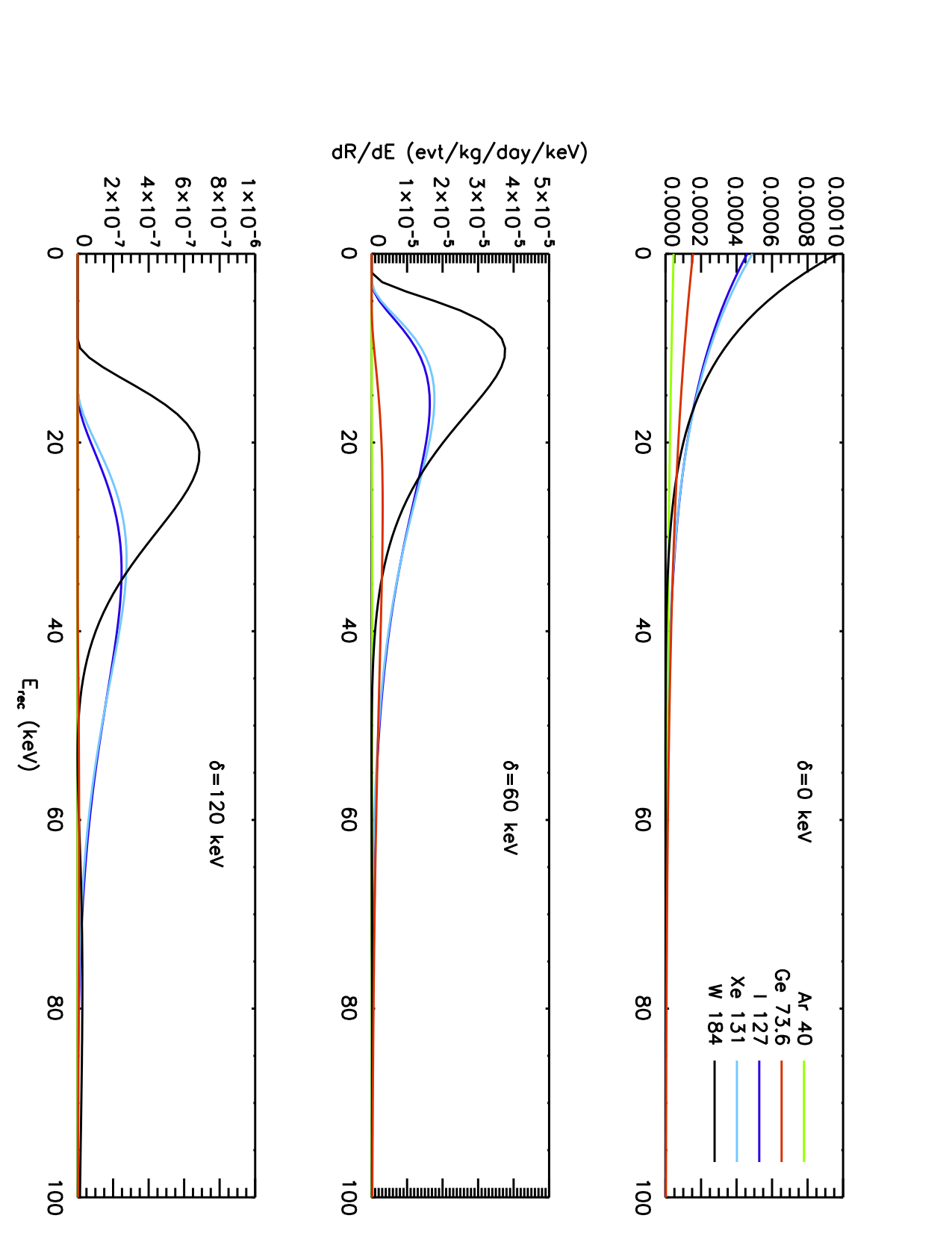}
\caption{Differential event rate for a WIMP scattering off different target nuclei: Ar (green), Ge (red), I (dark blue), Xe (fair blue) and W (black). The mass splitting values considered are: $\delta$=0~keV (top), 60~keV (middle) and 120~keV (bottom).}
\label{fig:rate_target}
\end{center}
\end{figure}

We have computed the differential event rate using the analytical solution of equation~(\ref{eq: WIMP rate}) proposed in \cite{SFG2006} and considering the same values as in the elastic case describe above. In figure~\ref{fig:rate_target} we show the effect of the the mass splitting $\delta$ in the event rate for different targets in the case of a WIMP mass of 100~GeV, a cross-section of 10$^{-8}$~pb, and $v_{\rm max}$=774~\kms. We observe that an increase in the mass splitting globally reduces the signal, even suppresses it at low recoil energies. Given the same $\delta$, heavier targets would be more sensitive to the signal. For instance an iodine target (which is the case of DAMA) could still detect a WIMP signal whereas a Ge detector would not be sensitive anymore.\\

\begin{figure}[h!]
\begin{center}
        \includegraphics[height=0.55\linewidth,angle=90]{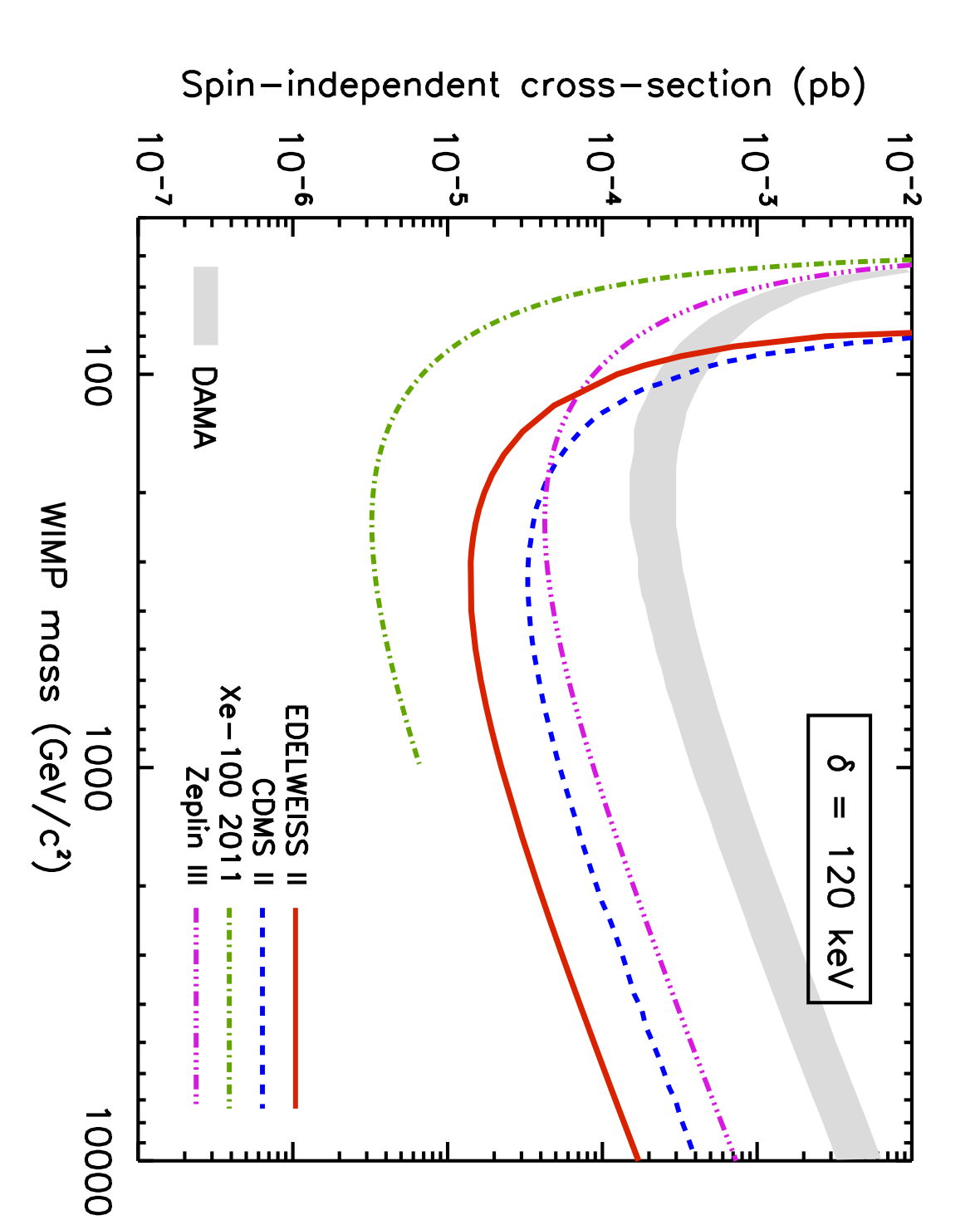}
\caption{WIMP-nucleon spin-independent cross-section upper limit at 90\% CL vs.~WIMP mass for inelastic scattering with $\delta$=120 keV. This analysis (solid red) together with the latest results from several direct detection experiments, and DAMA allowed region at 90\% CL.}
\label{fig:limit_idm120}
\end{center}
\end{figure}

We have interpreted the EDELWEISS-II results presented above in the inelastic scattering scenario. Following the same procedure as in the previous section, the WIMP-nucleon spin-independent cross-section upper limit has been computed for $\delta$=120 keV with the optimum interval method \cite{Ye2001}. The result is shown in figure~\ref{fig:limit_idm120} together with limits recently published by other direct detection experiments. The EDELWEISS-II limit excludes the DAMA region for WIMP masses greater than 90 GeV. The lack of events between 23 and 172 keV provides a very good sensitivity at large WIMP masses.

\section{Current status and prospects} \label{sec:current status prospects}
The EDELWEISS-II experiment has carried out a direct WIMP search with an array of ten 400-g Inter-Digit detectors, achieving an effective exposure of 384~\kgd. Five WIMP candidates in the energy range [20,200]~keV have been reported. The best sensitivity achieved in the elastic spin-independent WIMP-nucleon cross-section is 4.4$\times$10$^{-8}$~pb for a WIMP mass of 85~GeV. An interpretation of these results in the inelastic scattering scenario excludes the allowed DAMA region for WIMP masses greater than 90~GeV. The results are detailed in \cite{EDW2011}.\\

\begin{figure}[h!]
\begin{center}
 \begin{minipage}{0.49\linewidth}
       \includegraphics[width=\linewidth]{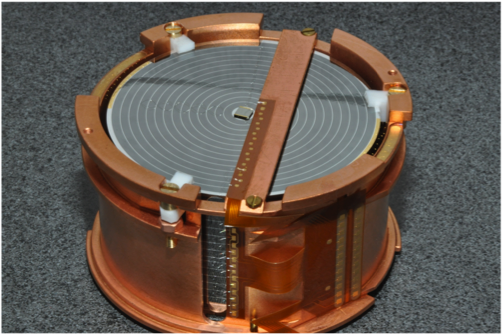}
\end{minipage}
\hfill   
\begin{minipage}{0.49\linewidth}
      \includegraphics[width=\linewidth]{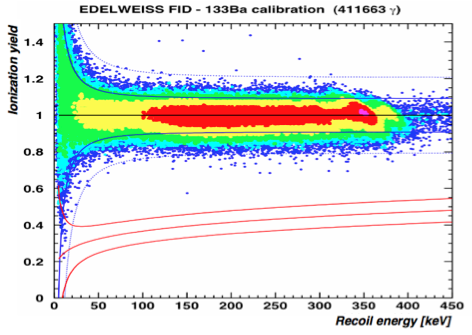}
\end{minipage}
\caption{Left: FID800 detector. Right: Gamma calibration of a FID800. The lines superimposed represent the average (resp. worst) 99.99\% rejection line for electron recoils in solid (resp. dashed) blue, and the 90\%-acceptance nuclear recoil band (red).}
\label{fig:FID800}
\end{center}
\end{figure}

The Inter-Digit detector technology has proven to be reliable and robust enough to perform direct detection of WIMPs at a competitive level. To go beyond the present performance, a new generation of detectors has been conceived: the Full Inter-Digit (FID). In this new design, the interleaved electrodes cover also the lateral surfaces of the bolometers and they are connected such that surface event rejection is possible in both fiducial and guard volumes. A first series of two 400-g and four 800-g FID detectors has been built and tested (see picture in figure~\ref{fig:FID800} (left)). The combination of an unprecedented mass of 800 g and the FID technology will significantly increase the fiducial mass of the detectors. In addition to this, the FID800 series will have two NTD sensors to have redundancy also in the heat measurement. Moreover, several surface treatments are under study to increase the surface event rejection.\\

Extensive $\gamma$ calibrations of FID800 detectors have already been performed at LSM. The ionisation yield vs.~recoil energy plot obtained is shown in figure~\ref{fig:FID800} (right). The lack of events in the nuclear recoil band with a statistics equivalent to that shown for ID detectors in figure~\ref{fig:neutron gamma calibration} (right) is proof of the improvement achieved.\\

Regarding the EDELWEISS-II setup, new upgrades in several parts (cryostat, shieldings, cabling, electronics) are foreseen to reduce the background and lower the energy threshold. The goal is to install 40 FID800 bolometers to reach a 3000~\kgd ~exposure in 2012 and a potential WIMP-nucleon cross-section sensitivity at the level of 5$\times$10$^{-9}$~pb.

The future European cryogenic experiment for direct detection of dark matter is the EURECA project, which intends to build an ultra-low background, ton-scale experiment combining different types of targets to reach cross-section values beyond 10$^{-9}$~pb. The installation site would be an extension of the present Laboratoire Souterrain de Modane.

\end{document}